\documentclass[prb,amsmath,amssymb,showpacs,preprint]{revtex4}

\usepackage{graphicx}
\def\be{\begin{equation}}
\def\ee{\end{equation}}
\def\ps{\tilde\psi}

\begin{document}

\title{Supercurrent fluctuations in short filaments}
\author{Jorge Berger}
\affiliation{Department of Physics, Ort Braude College, P. O. Box 78, 21982 Karmiel, Israel}
\email{jorge.berger@braude.ac.il}
\begin{abstract}
We evaluate the average and the standard deviation of the supercurrent in superconducting nanobridges, as functions of the temperature and the phase difference, in an equilibrium situation. We also evaluate the autocorrelation of the supercurrent as a function of the elapsed time. The behavior of supercurrent fluctuations is qualitatively different from from that of the normal current: they depend on the phase difference, have a different temperature dependence, and for appropriate range their standard deviation is independent of the probing time. We considered two radically different filaments and obtained very similar results for both. Fluctuations of the supercurrent can in principle be measured.
\end{abstract}
\maketitle

\section{Introduction}
Josephson junctions are of common use in many technological devices and their behavior as a circuit element is described in textbooks;\cite{Tin,Lik,Ba} the influence of thermal fluctuations on the normal current is described as Johnson noise. The voltage due to thermal noise was studied by Ambegaokar and Halperin.\cite{AH}

In this study we are interested in the equilibrium fluctuations of the supercurrent. Since supercurrent is actually an equilibrium variable rather than a diffusion process, we may expect---and will indeed find---that its fluctuations are qualitatively different from those of the normal current. Early works on fluctuations in junctions\cite{RS} state that there is no noise in supercurrent to quadratic order in the temperature. On the other hand, Averin and Imam\cite{Averin} found that supercurrent fluctuations in mesoscopic contacts are large on the scale of the classical shot noise.

The kind of junction considered in this article is a filament close to the critical temperature, which can be described by means of the Ginzburg--Landau model. If the entire filament has a critical temperature that is above the experimental temperature $T$, the filament may be regarded as a constriction; if parts of the filament have a critical temperature above $T$, it may be regarded as an SNS junction. In the case that fluctuations are ignored, the junction-like behavior of a filament has been studied for static\cite{Bara} and for dynamic\cite{Yako} situations.

The situation we will study here is a case of dynamic equilibrium. We will consider a superconducting filament that bridges between two ``banks." At the banks fluctuations are negligible and the order parameter will have fixed equilibrium values, whereas along the filament the order parameter and the electromagnetic potential fluctuate. In the absence of fluctuations, the current along the filament would be given by the current--phase relation.

\section{Method and Definitions}
We will use the time-dependent Ginzburg--Landau (TDGL) equations with Langevin terms, which will be handled numerically by means of finite differences. We have described this method in detail in the past\cite{Langevin,JPCM} and shown good agreement with statistical mechanics and with experiments.

For a 1D filament we define the gauge-invariant order parameter
$\ps (s)=\exp[(2\pi i/\Phi_0)\int_0^s A(s')ds']\psi (s)$, where $s$ is the arc length, $\psi$ the ``canonical" order parameter, $A$ the tangential component of the vector electromagnetic potential and $\Phi_0$ the quantum of flux. The 1D-TDGL equation can be written as\cite{giles}
\be
u\hbar \frac{\partial \ps}{\partial t}=-\left[\alpha +\beta |\ps |^2-\frac{\hbar^2}{2m}\frac{\partial^2}{\partial s^2}-\frac{1}{w}\frac{\partial w}{\partial s}\frac{\partial}{\partial s}\right]\ps \;,
\label{Gil}
\ee
where $m$ is the mass of a Cooper pair, $w(s)$ the cross section of the filament, and $u$, $\alpha $ and $\beta $ are material parameters; the sign of $\alpha $ determines whether the local critical temperature is above or below $T$.

We will take the boundary conditions $\ps (0,t)=\sqrt{-\alpha (0)/\beta (0)}$, $\ps (L,t)=\sqrt{-\alpha (L)/\beta (L)}\exp (i\gamma)$, where $L$ is the length of the filament; $\gamma $ is the gauge-invariant phase difference.

The scaled values of $\alpha $ and $\beta $ will be kept fixed in the present study. Since $\alpha $ and $\beta $ are functions of temperature, the physical meaning of this exposition is that for each temperature the experiment is performed on a different sample.

The normal current $I_N(s,t)$ and the supercurrent $I_S(s,t)$ are not separately constant along the filament. We define the supercurrent as the weighted average $I_S(t)=\int_0^L I_S(s,t)w(s)^{-1}ds/\int_0^L w(s)^{-1}ds$. $w(s)^{-1}$ appears as a natural weight in Refs. \onlinecite{Langevin,JPCM}; in this work we consider uniform cross section only, so that this weight and the last term in Eq.~(\ref{Gil}) can be ignored.

\section{Results}
We have examined two cases. One of them is a uniform filament with $\alpha =-\hbar ^2/mL^2$; in the second case we took $\alpha (x)=-(\hbar ^2/mL^2)\cos (2\pi x/L)$, so that the middle of this filament is nominally normal. In both cases we took 
$\beta =\hbar ^2 w/mL$ and the resistance of the filament as $\hbar /4ue^2$. The filament was divided into 30 computational cells and evolution was followed in steps of duration $1.3\times 10^{-4}umL^2/\hbar$. The first $10^7$ steps had the purpose of relaxation to typical values of $\ps$ and then $12\times 10^7$ steps were used for averaging. Our results do not depend appreciably on the resistance or the steps duration. The averages shown in the figures are averages over time and $I_{S0}$ denotes the supercurrent in the absence of thermal fluctuations.

Using BCS, dirty limit\cite{Tin} and free electron gas approximations, our choices for $\alpha $ and $\beta $ can be inverted and regarded as choices for the geometric parameters. We obtain
\be
L^2=\pi\hbar^2 k_F\ell_e/6mk_B(T_c-T) \;,\;\;\; w=0.51L/n_e\ell_e^2 \;,
\label{micro}
\ee
where $k_F$ is the Fermi wavevector, $\ell_e$ the mean free path, $T_c$ the critical temperature of the banks and $n_e$ the electron density.

\subsection{Uniform filament}

\begin{figure}
\scalebox{0.85}{\includegraphics{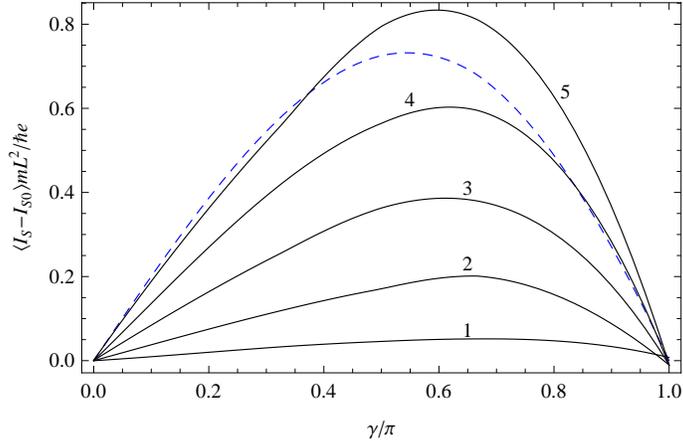}}%
\caption{\label{dcurr} Average deviation of the supercurrent from the value that would be obtained without thermal fluctuations, as a function of the phase difference and for several temperatures (solid curves). The temperature is $0.1n^2\hbar ^2/mL^2k_B$, where $n$ is the number marked next to each curve. For comparison, we have also drawn the curve $-I_{S0}/3$ (dashed line).}
\end{figure}

\begin{figure}
\scalebox{0.85}{\includegraphics{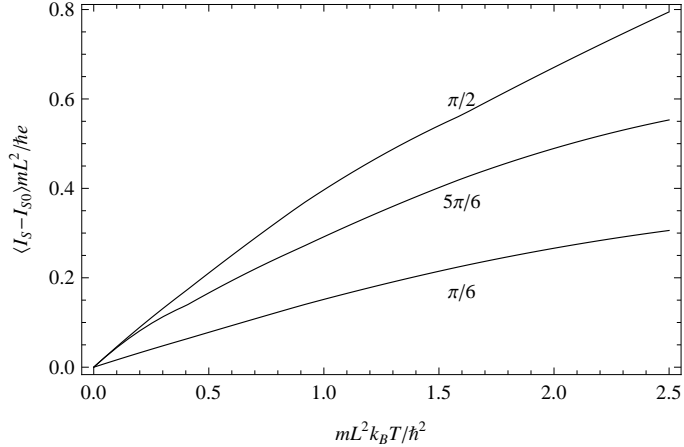}}%
\caption{\label{tdcurr} Deviation of the supercurrent from the fluctuation-free value, as a function of the temperature and for several phase differences, marked next to each curve. }
\end{figure}

Figure \ref{dcurr} shows the average deviation of $I_S$ from $I_{S0}$ as a function of $\gamma $ for several temperatures. We see that fluctuations not only lead to variance of the supercurrent, as in the case of normal current, but also lead to a shift of the average value. $I_S$ and $I_{S0}$ are negative for $0<\gamma <\pi$, so that $|\langle I_S\rangle |<|I_{S0}|$. For a material with critical temperature of the order of 1K, the highest temperature in Fig.~\ref{dcurr} corresponds to a filament length of the order of 30 nm and the highest current deviation, to the order of 10 nA. The $\gamma $-dependence of $\langle I_S\rangle -I_{S0}$ has some resemblance with that of $I_{S0}$, which has been included in the figure for comparison. Figure~\ref{tdcurr} presents $\langle I_S\rangle -I_{S0}$ as a function of temperature; it is apparent that this deviation behaves differently for different values of $\gamma $.

\begin{figure}
\scalebox{0.85}{\includegraphics{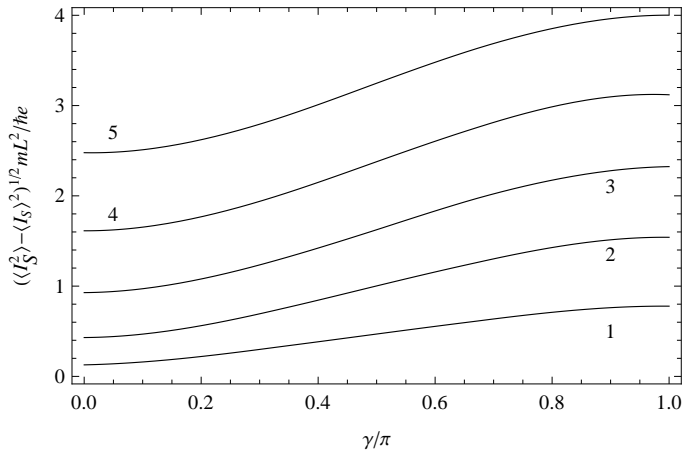}}%
\caption{\label{std} Standard deviation of the supercurrent, as a function of the phase difference, for several temperatures. The temperatures are the same as in Fig.~\ref{dcurr}.}
\end{figure}

Figure \ref{std} shows the standard deviation of $I_S$ as a function of $\gamma $. Note that whereas the standard deviation of Johnson noise is inversely proportional to the square root of the probing time (provided that it is long compared to $\hbar /k_BT$), the standard deviation of $I_S$ is independent of the probing time (provided that it is short compared to the decoherence time that we will find below). In addition, the standard deviation of $I_S$ depends on the phase difference, whereas that of $I_N$ does not. On the other hand, the present result does not support the scenario assumed in Ref.~\onlinecite{rectif}, according to which fluctuations of the order parameter just scale the supercurrent, while the shape of the current-phase relation remains fixed; if this were the case, the standard deviation of $I_S$ would be proportional to $I_{S0}$.

\begin{figure}
\scalebox{0.85}{\includegraphics{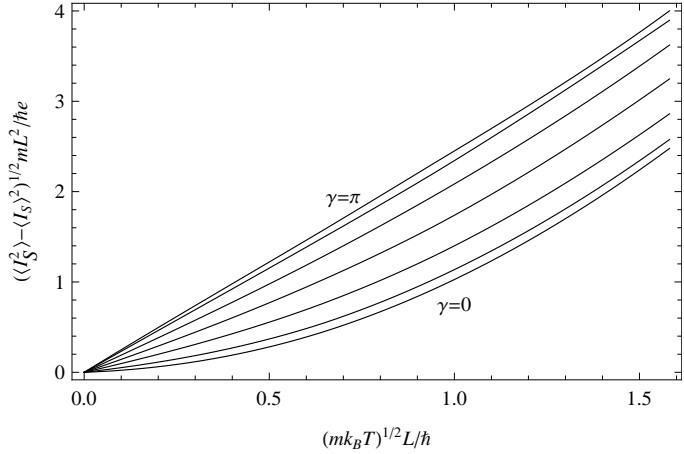}}%
\caption{\label{tstd} Temperature dependence of the standard deviation of the supercurrent. The lowest curve is for $\gamma =0$, the highest for $\gamma =\pi$, and the curves in between are in steps of $\pi /6$.}
\end{figure}

Figure \ref{tstd} shows the temperature dependence of the standard deviation of $I_S$. Whereas for the normal current the standard deviation is proportional to $T^{1/2}$, for the supercurrent this scaling occurs only in the case $\gamma =\pi$; as $\gamma$ decreases towards 0, the initial slope of the curves decreases.

\begin{figure}
\scalebox{0.85}{\includegraphics{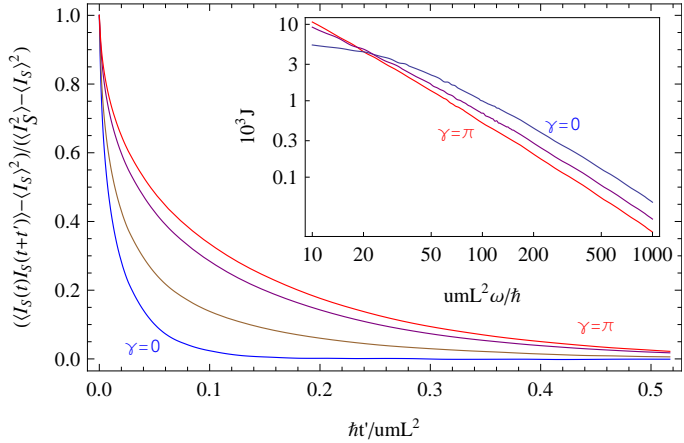}}%
\caption{\label{correl} Autocorrelation function of $I_S$, as a function of the elapsed time $t'$, for temperature $2.5\hbar ^2/mL^2k_B$ and $\gamma =n\pi /3$, $n=0,1,2,3$. The other parameters are as in Fig.~\ref{dcurr}. Inset: spectral density for $\gamma =0,\pi /2,\pi$.}
\end{figure}

Figure \ref{correl} shows the autocorrelation of the supercurrent, $K(t')=(\langle I_S(t+t')I_S(t)\rangle -\langle I_S\rangle^2)/(\langle I_S^2\rangle -\langle I_S\rangle^2)$ for $0\le\gamma\le\pi$. We see that, the smaller the value of $\gamma $, the shorter the typical time required to ``forget" a previous value of $I_S$. The inset shows the spectral density $J(\omega )=(1/\pi )\int_0^\infty K(t)\cos(\omega t)dt$.

If $I_S$ is measured many times during probing periods of length $\tau$ and $K(\tau )\sim 1$, then each measurement can essentially be regarded as instantaneous and the standard deviation of $I_S$ is given by Fig.~\ref{std}, with practically no $\tau$-dependence; if $K(\tau )\sim 0$, then fluctuations of $I_S$ essentially become white noise and the standard deviation of $I_S$ should decrease as $\tau^{-1/2}$. According to Fig.~\ref{correl}, the crossover value of $\tau$ (the ``decoherence time") increases with $\gamma $ and is of the order of $0.1umL^2/\hbar$ (for $L\sim 10^{-7}$m, this is of the order of $10^{-10}$s).

\begin{figure}
\scalebox{0.85}{\includegraphics{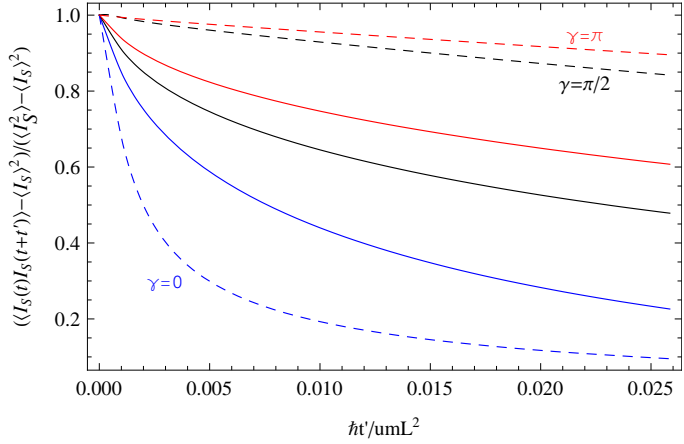}}%
\caption{\label{nofl} Influence of temperature on the autocorrelation function. The dashed lines are for temperature close to zero and the continuous lines for $2.5\hbar ^2/mL^2k_B$. The lower (blue online) lines are for $\gamma =0$, the middle lines for $\gamma =\pi/2$, and the upper (red online) lines for $\gamma =\pi$.}
\end{figure}

The faster relaxation of fluctuations for small $\gamma $ is counterintuitive, since according to Fig.~\ref{std} the influence of thermal agitation is stronger for large $\gamma $, and we might expect this agitation to destroy any particular configuration of the order parameter that results in a particular supercurrent at a given time. Figure~\ref{nofl} shows that the $\gamma $-dependence of the relaxation times of $I_S$ is not dominated by thermal agitation, but rather by the fluctuationless dynamics. At $T\sim 0$ memory loss of particular configurations is faster for small $\gamma $; thermal agitation moderates the $\gamma $-dependence. It is interesting to note that in the case $\gamma =0$ thermal agitation leads to delay of the relaxation.

The results in Figs. \ref{std} and \ref{nofl} can be interpreted as follows. In the case $\gamma =0$ the order parameter is pinned with the same phase at both boundaries, giving a large energy advantage to a uniform order parameter all along the filament. Therefore, the system is comparatively rigid: deviations from the equilibrium configuration are small and return to equilibrium is fast. In the case $\gamma =\pi$ the order parameter is pinned with opposite phases at the boundaries, so that the system is frustrated and comparatively indifferent; as a consequence, deviations from equilibrium are large and return to it is slow.

\subsection{SNS junction}
Here we report the results for the case $\alpha =-(\hbar ^2/mL^2)\cos (2\pi x/L)$. Although taking a value of $\alpha $ that vanishes on the average may look as a drastic change, the results are remarkably similar to those of the previous section.

\begin{figure}
\scalebox{0.85}{\includegraphics{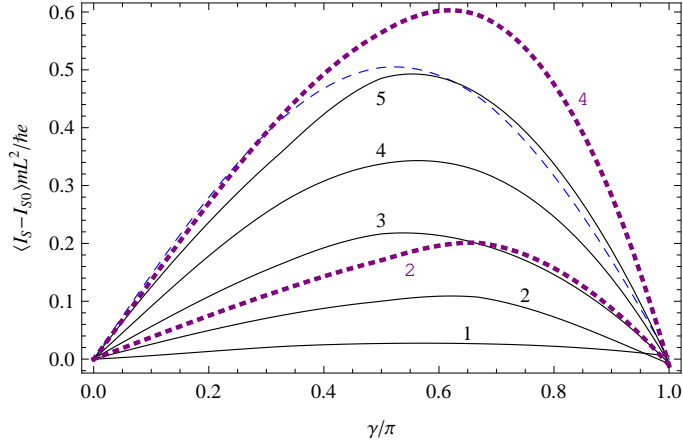}}%
\caption{\label{dcurrN} Reduction of average supercurrent due to thermal fluctuations in the case $\alpha =-(\hbar ^2/mL^2)\cos (2\pi x/L)$. Annotations as in Fig.~\ref{dcurr}. For comparison, the curves for $\alpha =-\hbar ^2/mL^2$, $T=0.4\hbar ^2/mL^2k_B$ and $T=1.6\hbar ^2/mL^2k_B$
have been included as dotted lines.}
\end{figure}

Figure \ref{dcurrN} shows the average deviation of $I_S(\gamma )$ from $I_{S0}(\gamma )$ for the same temperatures as in Fig.~\ref{dcurr}. We note that in the present case the reduction of $I_S$ is smaller than in the case of a uniform filament. This behavior makes sense, since when the entire filament is superconducting fluctuations may only be expected to destroy superconductivity; on the other hand, when half of the filament is normal and supercurrent can only be expected due to proximity or fluctuations, fluctuations also have a supportive effect. For $\gamma =\pi$ the average current has to vanish by symmetry; the scattering in this value serves as a measure of the accuracy of our results.

\begin{figure}
\scalebox{0.85}{\includegraphics{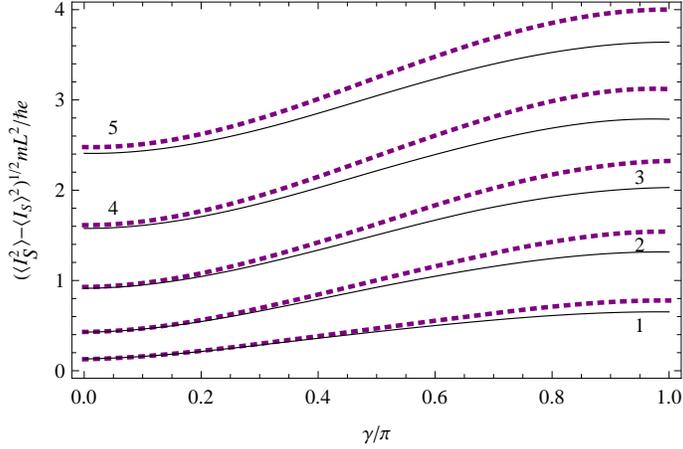}}%
\caption{\label{stdN} Standard deviation of the supercurrent in the case $\alpha =-(\hbar ^2/mL^2)\cos (2\pi x/L)$, for several temperatures. The temperatures are the same as in Fig.~\ref{dcurr}. For comparison, the curves of Fig.~\ref{std} are redrawn as dotted lines.}
\end{figure}

\begin{figure}
\scalebox{0.85}{\includegraphics{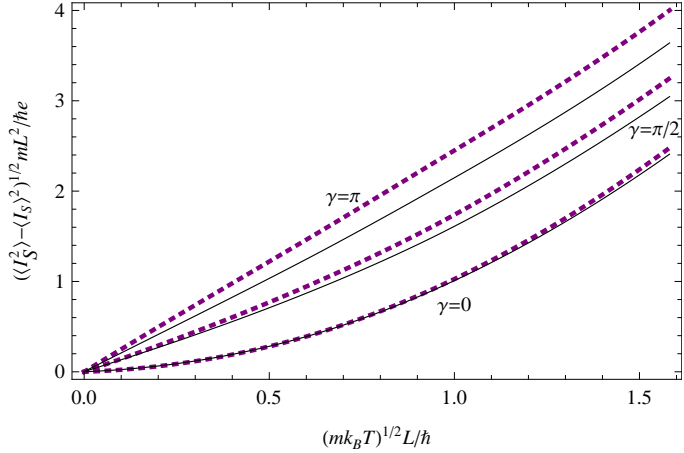}}%
\caption{\label{tstdN} Temperature dependence of the standard deviation of the supercurrent. Solid lines: $\alpha =-(\hbar ^2/mL^2)\cos (2\pi x/L)$; dotted: $\alpha =-\hbar ^2/mL^2$, redrawn from Fig.~\ref{tstd}.}
\end{figure}

Figure \ref{stdN} shows the standard deviation of $I_S(\gamma )$ and compares it with that of a uniform filament. For small values of $\gamma $ the standard deviation is similar to that of Fig.~\ref{std}, but as $\gamma $ approaches $\pi$ the standard deviation in the present case is noticeable smaller than that of a uniform filament. This result may seem surprising, since in the present case superconductivity is more fragile and we might expect to larger fluctuations. A possible explanation could be that when $\alpha $ and $\beta $ are both positive, large fluctuations of the order parameter in the middle of the filament are inhibited, leading to smaller fluctuations of $I_S$. Figure \ref{tstdN} presents these results as functions of the temperature. In the present case the standard deviation does not become proportional to $T^{1/2}$ even for $\gamma =\pi$.

\begin{figure}
\scalebox{0.85}{\includegraphics{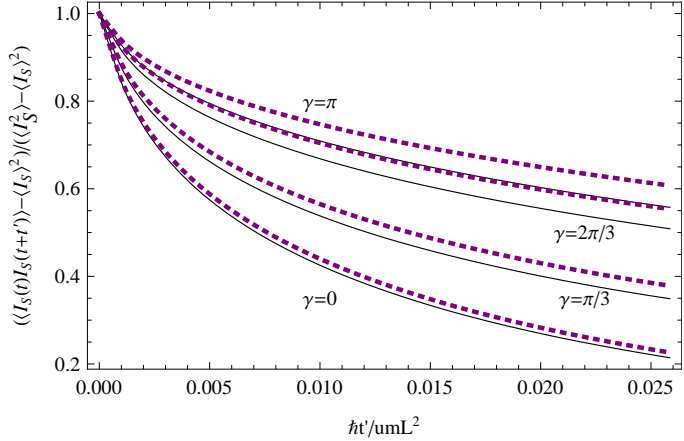}}%
\caption{\label{startN} Autocorrelation of the supercurrent, as a function of the elapsed time, for $\alpha =-(\hbar ^2/mL^2)\cos (2\pi x/L)$ and several values of $\gamma$. The other parameters are as in Fig.~\ref{correl}. The dotted lines are for $\alpha =-\hbar ^2/mL^2$.}
\end{figure} 

Figure \ref{startN} compares the autocorrelation functions for the cases $\alpha (x)=-(\hbar ^2/mL^2)\cos (2\pi x/L)$ and $\alpha =-\hbar ^2/mL^2$. Again, we find that the difference is remarkably small, especially for small $\gamma$. Weakening of superconductivity leads to a slightly faster decoherence.

\section{Discussion}
We have evaluated numerically the thermal fluctuations of the supercurrent along filaments that bridge between two banks (superconducting pieces with dimensions such that fluctuations in them are negligible) in an equilibrium situation. One case we considered was that of a uniform filament, made of the same material as the banks; the second case was that of a filament equal to the first at the contact points, but normal in the middle.

These fluctuations have non trivial properties, qualitatively different from those of normal current. On the other hand, in spite of the blatant difference between both considered filaments, the difference between the fluctuations in them is minor, suggesting that the results we have found are generic.

Experimentally, a phase difference may be  applied by connecting the banks so that together with the filament they become a closed circuit that encloses a known magnetic flux; the current can  then be sensed through the field that it induces. Measurement of $I_{S0}$ is complicated by the fact that changing the temperature would also change the value of $\alpha $.

Let us now discuss measurement of the standard deviation of $I_S$. The probing time should be of the order of $mL^2/\hbar $, which in view of Eq.~(\ref{micro}) is of the order of $\hbar k_F\ell_e/k_B(T_c-T)$, and ought to be accessible for temperature sufficiently close to $T_c$. For this probing time, using our results in Fig.\ \ref{tstd} or \ref{tstdN}, Eq.~(\ref{micro}), and the free electron gas expression for the resistivity, we estimate that the ratio between the standard deviation of $I_S$ and that of Johnson noise is of the order of $(\ell_e k_F)^{1/2}$; since this is not a small number, fluctuations of the supercurrent are not obscured by those of the normal current and could in principle be observed.

The present study may be regarded as a feasibility test for the influence of supercurrent fluctuations. 
Analytic treatments that uncover the scalings and asymptotic relations, as well as the non equilibrium behavior, are still required.  

\begin{acknowledgments}
This research was supported by the Israel Science Foundation, grant No.\ 249/10. I have benefited from comments from Grzegorz Jung and Baruch Rosenstein.
\end{acknowledgments}
 
\appendix
\section{Evaluation of $\langle I_S\rangle$ using a transfer operator technique}
The equilibrium average of the supercurrent is given by 
\be
\langle I_S\rangle=\langle I_S+I_N\rangle =\frac{2\pi k_BT}{\Phi_0}\frac{\partial \ln Z}{\partial\gamma} \;, 
\label{current} \ee
where $Z$ is the partition function, which has to be derived from the Ginzburg--Landau free energy
\be
F=\int_0^L w ds [\alpha |\ps |^2+(\beta/2) |\ps |^4+(\hbar^2/2m)|\partial\ps /\partial s|^2] \;.
\ee
For a uniform filament and $\alpha <0$, writing $s=Lt$ and $({\rm Re}\ps ,{\rm Im}\ps )=\sqrt{-\alpha /\beta }\,{\bf r}$, $F$ becomes
\be
F=-\frac{w\alpha \hbar^2}{L\beta m}\int_{0}^1 dt \left( \frac{1}{2}\left|
\frac{d{\bf r}}{dt }\right|^2+ V\right) \;, 
\label{F} \ee
with
\be
V=(\alpha L^2m/\hbar^2)(r^2-r^4/2)\;. 
\label{V0} \ee
Following Ref.~\onlinecite{Scal}, a function ${\bf r}(t)$ is interpreted as a microstate of the system and $F$ as
the energy of the system for that microstate. It follows that the partition function is 
\be
Z=C\int {\cal D}{\bf r}\exp (-F/k_B T) \;,
\label{Z1} \ee 
where $\int {\cal D}{\bf r}$ denotes integration over all functions ${\bf r}(t)$ with boundary conditions ${\bf r}(0)={\bf a}_0=(1,0)$ and ${\bf r}(1)={\bf a}_1=(\cos\gamma ,\sin\gamma )$, and $C$ is an irrelevant multiplicative constant. Dividing the integral in Eq.~(\ref{F}) into $N$ segments, $N\gg 1$, and introducing into Eq.~(\ref{Z1}) we can write in this limit
\be
Z=\left(\frac{N}{2\pi S}\right)^{N-1}\int d{\bf r}_1\dots d{\bf r}_N\delta ({\bf r}_1-{\bf a}_0)\delta ({\bf r}_N-{\bf a}_1)
\exp[-\frac{f({\bf r}_{N},{\bf r}_{N-1})}{NS}]\dots\exp[-\frac{f({\bf r}_2,{\bf r}_1)}{NS}] \;,
\ee
with $f({\bf r}_{i+1},{\bf r}_i)=V({\bf r}_{i+1})+(N^2/2)|{\bf r}_{i+1}-{\bf r}_i|^2$, $S=-k_BTL\beta m/w\alpha \hbar^2$, and the prefactor has been chosen for convenience.

Noting that $\delta ({\bf r}-{\bf r}')=\sum_n\Psi^*_n({\bf r})\Psi_n({\bf r}')$, where $\{\Psi_n\}$ is any complete set of normalized eigenstates, $Z$ becomes
\begin{eqnarray}
Z&=&\sum_{n,n'}\Psi_{n'}({\bf a}_1)\Psi^*_n({\bf a}_0)\int d{\bf r}_N \Psi^*_{n'}({\bf r}_N)\cdots 
\int d{\bf r}_i (N/2\pi S)\exp[-f({\bf r}_{i+1},{\bf r}_i)/NS] \cdots\nonumber \\
&\times& \int d{\bf r}_1 (N/2\pi S)\exp[-f({\bf r}_2,{\bf r}_1)/NS]\Psi_n({\bf r}_1)\;.
\end{eqnarray}
If the $\Psi_n$ are chosen so that they obey the eigenvalue equation 
\be
\int d{\bf r}_i (N/2\pi S)\exp[-f({\bf r}_{i+1},{\bf r}_i)/NS]\Psi_n({\bf r}_i)=\exp[-\epsilon_n/NS]\Psi_n({\bf r}_{i+1}) \;, 
\label{defeigs}
\ee
then
\be
Z=\sum_n\Psi_n({\bf a}_1)\Psi^*_n({\bf a}_0)\exp(-\epsilon_n/S)\;.
\label{ZPsi}
\ee

Expanding $\Psi_n({\bf r}_i)$ in powers of ${\bf r}_i-{\bf r}_{i+1}$ around $\Psi_n({\bf r}_{i+1})$, which is equivalent to an expansion in powers of $N^{-1/2}$, the integral in Eq.~(\ref{defeigs}) can be performed and $\Psi_n$ is found to obey the eigenvalue equation\cite{Scal}
\be
[-(S^2/2)\mathbf{\nabla }^2+V]\Psi_n =\epsilon_n\Psi_n \;,
\ee
where $\mathbf{\nabla }^2$ is the Laplacian with respect to ${\bf r}$.

In polar coordinates ${\bf r}=r(\cos\theta,\sin\theta)$, the angular momentum operator is $L_z=-i\partial /\partial \theta $. Noting that\cite{Oppen} $L_z$ commutes with the ``Hamiltonian" $[-(S^2/2)\mathbf{\nabla }^2+V]$, the $\Psi_n$ can be chosen so that they are also eigenstates of $L_z$, i.e.\ they can have the form $\Psi_{n,\ell}[r(\cos\theta,\sin\theta)]=R_{n,\ell}(r)\exp (i\ell\theta )$, where $R_{n,\ell}$ is a real function and $\ell$ an integer. Introducing this form into Eq.~(\ref{ZPsi}) we obtain
\be 
Z=\sum_\ell\exp (i\ell\gamma )Z_\ell \;, 
\label{Z3} \ee with 
\be Z_\ell =\sum_n R^2_{n,\ell}(1)\exp(-\epsilon_{n,\ell}/S) \;, \label{ZL} \ee 
where summation in Eq.~(\ref{Z3}) is made over all integers and in
Eq.~(\ref{ZL}) over all the states with total angular momentum $\ell$.

Since the Hamiltonian is symmetric under the transformation ${\bf r}\rightarrow -{\bf r}$, we can also write
\be Z=Z_0+2\sum_{\ell =1}^\infty\cos (\ell\gamma )Z_\ell \;.
\label{Z4} \ee
Applying Eq.~(\ref{current}) we obtain
\be
\langle I_S\rangle =-\frac{4\pi k_BT}{\Phi_0 Z}\sum_{\ell =1}^\infty\sin (\ell\gamma )\ell Z_\ell \;. 
\label{current1} \ee

In order to complete the evaluation, Eqs.\ (\ref{current1}) and (\ref{ZL}) have to be supplemented with the values of $\epsilon_{n,\ell}$ and $R_{n,\ell}(1)$. Since only for small values of $\epsilon_{n,\ell}$ there is an appreciable contribution, this is accomplished as follows. We start from a basis Hamiltonian $H_B=-(S^2/2)\mathbf{\nabla}^2+(k^2/2)r^2$, which has known eigenfunctions 
$R^B_{n,\ell}(r)=C_{n,\ell}r^{|\ell |}e^{-kr^2/2S}\,_1F_1(-n,|\ell |+1,kr^2/S)$ and eigenvalues $\epsilon^B_{n,\ell}=Sk(2n+|\ell|+1)$, where
$C_{n,\ell}$ is the normalization constant and $_1F_1$ is Kummer's hypergeometric function. $k$ is still a free parameter. We select a subspace of the Hilbert space which has a moderate number of low-energy eigenstates of $H_B$ as a basis, project the true Hamiltonian into this subspace, and then diagonalize this truncated Hamiltonian. 

The value of $k$ is chosen as follows. If the Hamiltonian were not truncated, its lowest eigenvalue would be independent of $k$; for the truncated Hamiltonian and for a given value $\ell$, the lowest eigenvalue is effectively independent of $k$ within a certain range and rises beyond this range. We choose $k(\ell)$ in the middle of this range.

\begin{figure}
\scalebox{0.85}{\includegraphics{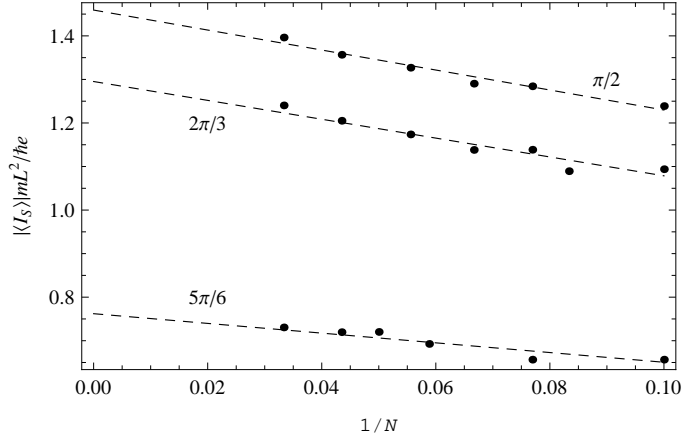}}%
\caption{\label{app} Supercurrent as a function of the length of the computational segments. The dashed lines extrapolate to the values obtained by the method described in this appendix. The temperature is $2.5\hbar ^2/mL^2k_B$, the phase difference is marked next to each line and the other parameters are as in Fig.~\ref{dcurr}. We kept 12 terms in expansion (\ref{current1}).}
\end{figure}

Figure \ref{app} compares between the computational method taken from Refs.\ \onlinecite{Langevin,JPCM} and the method developed in this appendix. The comparison requires extrapolation to the continuum limit, $N\rightarrow \infty $. The convergence of $\langle I_S\rangle$ to the continuum limit is rather slow, but the variance of $I_S$ seems to saturate for $N\sim 20$. For temperatures lower than $0.4\hbar ^2/mL^2k_B$ convergence in Eq.~(\ref{current1}) becomes slow.

\end{document}